\documentclass[manuscript,screen]{acmart}

\AtBeginDocument{%
  \providecommand\BibTeX{{%
    \normalfont B\kern-0.5em{\scshape i\kern-0.25em b}\kern-0.8em\TeX}}}

\setcopyright{acmlicensed}
\copyrightyear{2018}
\acmYear{2018}
\acmDOI{XXXXXXX.XXXXXXX}

\acmConference[Conference acronym 'XX]{Make sure to enter the correct
  conference title from your rights confirmation emai}{June 03--05,
  2018}{Woodstock, NY}
\acmISBN{978-1-4503-XXXX-X/18/06}

\definecolor{tableline-gray}{gray}{0.9}
\usepackage{float}
\usepackage{pifont}
\usepackage{balance}  
\usepackage{multirow}
\usepackage{graphicx}
\usepackage{dcolumn}
\usepackage{colortbl}
\usepackage[T1]{fontenc}
\usepackage{subcaption}
\usepackage{makecell}
\usepackage{bm}
\usepackage{fixme}
\usepackage{hyperref}
\usepackage{hhline}
\usepackage{csquotes}
\usepackage{changepage} 
\usepackage{xcolor}
\usepackage{tikz}
\usetikzlibrary{mindmap, arrows.meta, positioning}
\usepackage{booktabs}
\usepackage{caption}
\usepackage{float}

\begin{document}

\title{Insights in Adaptation: Examining Self-reflection Strategies of Job Seekers with Visual Impairments in India}

\author{Akshay Kolgar Nayak}
\orcid{0009-0000-9992-9706}
\affiliation{%
  \institution{Old Dominion University}
  \department{Department of Computer Science}
  \city{Norfolk}
  \state{Virginia}
  \country{USA}
}
\email{anaya001@odu.edu}

\author{Yash Prakash}
\orcid{0000-0001-8593-327X}
\affiliation{%
  \institution{Old Dominion University}
  \department{Department of Computer Science}
  \city{Norfolk}
  \state{Virginia}
  \country{USA}
}
\email{yprak001@odu.edu}

\author{Sampath Jayarathna}
\orcid{0000-0002-4879-7309}
\affiliation{%
  \institution{Old Dominion University}
  \department{Department of Computer Science}
  \city{Norfolk}
  \state{Virginia}
  \country{USA}
}
\email{sampath@cs.odu.edu}

\author{Hae-Na Lee}
\orcid{0000-0002-2183-1722}
\affiliation{%
  \institution{Michigan State University}
  \department{Department of Computer Science and Engineering}
  \city{East Lansing}
  \state{Michigan}
  \country{USA}
}
\email{leehaena@msu.edu}

\author{Vikas Ashok}
\orcid{0000-0002-4772-1265}
\affiliation{%
  \institution{Old Dominion University}
  \department{Department of Computer Science}
  \city{Norfolk}
  \state{Virginia}
  \country{USA}
}
\email{vganjigu@odu.edu}

\renewcommand{\shortauthors}{Nayak, et al.}

\begin{abstract}
Significant changes in the digital employment landscape, driven by rapid technological advancements and the COVID-19 pandemic, have introduced new opportunities for blind and visually impaired (BVI) individuals in developing countries like India. However, a significant portion of the BVI population in India remains unemployed despite extensive accessibility advancements and job search interventions. Therefore, we conducted semi-structured interviews with 20 BVI persons who were either pursuing or recently sought employment in the digital industry. Our findings reveal that despite gaining digital literacy and extensive training, BVI individuals struggle to meet industry requirements for fulfilling job openings. While they engage in self-reflection to identify shortcomings in their approach and skills, they lack constructive feedback from peers and recruiters. Moreover, the numerous job intervention tools are limited in their ability to meet the unique needs of BVI job seekers. Our results therefore provide key insights that inform the design of future collaborative intervention systems that offer personalized feedback for BVI individuals, effectively guiding their self-reflection process and subsequent job search behaviors, and potentially leading to improved employment outcomes.
\end{abstract} 




\maketitle
\section{Introduction}
Job-seeking has evolved into a dynamic, self-regulated, and socially-driven process, characterized by a series of interconnected activities designed to secure employment~\cite{bainbridge2018job, wanberg2020job}. This transformation has been propelled by advancements in technology, the proliferation of social media platforms, and growing recognition of the importance of networking and shared knowledge in professional development. Individuals pursue employment at various life stages to address financial needs, engage in social interactions, foster personal development, and achieve career growth~\cite{furnham1983employment, wheeler2018navigating}. The job-seeking process typically follows a sequential progression~\cite{blau1994testing, saks2000change}, encompassing a `preparatory phase' and an `active phase'. During the preparatory phase, job seekers engage in planning and gathering information about job opportunities from a multitude of sources, often collaborating with peers, mentors, and professional networks to refine their skills and identify suitable career paths. The active phase involves crafting applications, attending interviews, and directly interacting with potential employers, where the ability to demonstrate developed skills and adaptability becomes crucial. Throughout the whole process, individuals continuously interact with others, seeking their advice to refine their skills and gain insights to enhance their readiness for the job market. However, this multifaceted and socially dynamic nature of job seeking, which relies heavily on digital platforms and collaborative interactions, can present unique challenges for blind and visually impaired (BVI) individuals~\cite{bainbridge2018job, gupta2021employment}.

In the year 2020, global estimates indicated that approximately $338$ million people were dealing with vision impairments~\cite{bourne2021trends}, a number slightly exceeding the entire population of the United States ($334.9$ million in 2023). Within this group, $43.3$ million individuals were categorized as blind, and an additional $295$ million were estimated to have moderate to severe vision impairment, i.e., low vision. India, in particular, has one of the largest populations of people with visual impairments, with approximately $70$ million vision-impaired persons and $4.95$ million blind individuals~\cite{sheetvisual,kameswaran2023advocacy}. Vision impairment in such developing countries, coupled with inaccessible education~\cite{vashistha2014educational} and limited support for accessibility in workplaces~\cite{pal2012assistive}, not only affects personal well-being but also has broader socioeconomic implications, as it often leads to increased poverty due to reduced employment opportunities and productivity~\cite{burton2021lancet,langelaan2007impact,mojon2010impact}. Unfortunately, despite several government initiatives and measures as well as the availability of non-profit training services, only about $100,000$ BVI individuals in India have succeeded in gaining employment in industries so far~\cite{blindreport2023}, leaving a significant percentage of the country's BVI population unemployed.

India has rapidly emerged as a central hub for offshore projects, driven by several key factors, including the increasing availability of high-speed internet, the rise of e-commerce, government initiatives, and the COVID-19 pandemic~\cite{financialexpress,toireport}. This shift has created opportunities in remote work, freelance gigs, and online platform-based jobs. However, within the `digital workspace', there exists a prevalent belief that BVI individuals are unable to compete or collaborate effectively with sighted peers~\cite{basu2023barriers}. This perception is primarily attributed to the numerous barriers the BVIs may encounter at work~\cite{golub2006model,basu2023barriers,cha2024understanding}. Prior research has highlighted such barriers encountered by BVI individuals during their employment search process across various domains in India. These include employer prejudice or adverse perceptions, communication difficulties, and issues related to lack of content accessibility~\cite{AIF2015JobMapping,blindwelfaresociety2024,carter2021preparing}. In an attempt to address these barriers, prior works have proposed facilitators that aid BVIs in job search including harnessing personal networks, establishing clear goals with the assistance from others, accessing information tailored to their needs, and obtaining job-search training~\cite{liu2014effectiveness, silverman2019understanding}. However, a critical knowledge gap remains in understanding how these barriers influence the job-search behaviors of BVI individuals in the first place. Specifically, little is known about how they evaluate and reflect on their skills through collective feedback, perceive societal acceptance, and respond to negative experiences while maintaining motivation and self-efficacy throughout the process. Furthermore, the adequacy of existing employment intervention systems in addressing their unique needs, both as individuals with disabilities and as members of a marginalized group in developing societies, requires deeper investigation.

We strive to fill this knowledge gap by retrospectively examining the job search experiences of BVI individuals in India, focusing on the educational, technological, and peer support available to guide them towards positive and effective self-reflection on their abilities and the needs of the industry. Our study specifically seeks to address the following research questions:

\begin{itemize}
    \item \textbf{RQ1: } How do societal factors such as social bias, collective knowledge, and community expectations influence the job-seeking experiences of BVI individuals in developing countries like India?
    \item \textbf{RQ2: } How do peer interactions and interview feedback influence self-reflection and subsequent career decision-making among BVI job seekers?
    \item \textbf{RQ3: } What are the strengths and weaknesses of the current job-intervention tools towards facilitating productive self-reflection among BVI job seekers?

\end{itemize}

To answers these research questions, we conducted an interview study\footnote{This is a standard procedure in CSCW research, particularly when involving individuals from ability-diverse and marginalized groups~\cite{das2019doesn,kameswaran2023advocacy}.} with $20$ BVI participants in India who have been actively engaged in the Information Technology (IT) job search and interview process at some point in their professional careers. Our analysis revealed key insights into the job search behavior of BVI individuals, notably:
(i) BVI individuals often begin their job search with high self-efficacy, which may lead them to overlook the practical gaps between their skills and the industry's demands.; (ii) BVI individuals feel that they need to put additional efforts to convince recruiters of their digital competence and ability to execute tasks without assistance; (iii) BVI job seekers are highly dependent on explicit feedback to identify areas for improvement and skill development, but such feedback is often limited or too generic both in the interviews and in the societies they grew up in.; (iv) Participants rarely consider self-reflection to be an isolated process. Instead, they view it as a social process involving peer feedback and guidance from mentors; and (v) BVI individuals strongly believe that current job intervention tools, including human tutoring, are not tailored for addressing their unique needs.

In sum, this study makes the following contributions to the CSCW literature:
\begin{itemize}
    \item We extend prior CSCW research on job-seeking experiences of BVI individuals beyond the context of developed nations (Global North), focusing on the unique challenges faced by BVI job seekers within a linguistically diverse, economically stratified, and socially complex developing societies like that in India (Global South).
    \item Building upon the broad literature on the significance of self-reflection in promoting continuous learning and skill development among underrepresented job seekers, we examine how this process manifests among BVI individuals through peer interactions and collective knowledge sharing.
    \item We uncover the capabilities and limitations of existing job-intervention systems towards fostering a positive self-reflective process among BVI job seekers, and provide design suggestions for intelligent job-intervention systems that can accommodate their unique needs and prepare them for real-world collaborative work environments.
\end{itemize}

\section{Background and Related Work}

\subsection{Self-reflection for Skill Development}\label{RW1}
Self-reflection has been defined as the process a person engages in to retrospectively examine their past learning experiences and the actions they undertook to facilitate learning (i.e., self-reflection on how learning occurred), alongside the investigation of links between the imparted knowledge and the person's conceptions of it (i.e., self-reflection on the acquired knowledge)~\cite{lew2011self,dewey2022we,mann2009reflection,boud2013reflection}.
Previous research has extensively explored this self-regulated process as a means to promote continuous learning, skill development, and elevated self-efficacy among diverse groups of people, especially those who actively seek employment opportunities at different stages in their life~\cite{furnham1983employment,wheeler2018navigating,kanfer2001job,van2013moving,granovetter1973extend,van2021job,wanberg2010job,van201812,noordzij2013effects,liu2014self,da2018insight,creed2009goal,burnette2013mind,bohlmeijer2003effects,mols2016technologies,lackner2017helping, siebert2013reflection,wanberg2012navigating}. For instance, an early study conducted with employed BVI individuals in the US highlighted how they must continuously evaluate their abilities, identify areas requiring support or skill enhancement, and align their strengths with job opportunities to achieve successful employment outcomes~\cite{crudden1998comprehensive}. This self-reflective process also enabled them to determine the specific accommodations or adaptive technologies necessary for effective workplace performance~\cite{crudden1998comprehensive,dillahunt2017uncovering,dillahunt2020positive}.

Self-reflection is inherently a social and collaborative process, shaped by interactions with coworkers, mentors, and peers who provide feedback, support, and role modeling~\cite{studd2024exploring,clarke2007reflective,taylor2020designing}. Positive reinforcement from inclusive employers and advocacy groups have further empowered BVI individuals to view themselves as capable contributors, fostering adaptation and growth ~\cite{Kornbluh2021,Akari2020,AFB2023}. Conversely, negative attitudes or systemic barriers have often prompted them to reassess their strategies and seek alternative career pathways through supportive networks~\cite{crudden1998comprehensive,crudden1999barriers,basu2023barriers}. While these findings are broadly applicable across contexts and time, the specific experiences of BVI job-seekers in the resource-constrained Global South, where access to accessible education and societal understanding of their capabilities are minimal, remain under-explored~\cite{richardson2019geographies,uk2015high}.

Considering the importance of this self-reflective process among underrepresented job-seekers, researchers have also proposed intervention strategies and tools that empower these job-seekers to self-reflect and validate skills relevant to the current demands of the industry~\cite{siebert2013reflection}. These tools aim to develop cognitive skills, such as learning strategies, problem-solving, and critical thinking, as well as meta-cognitive skills, which include planning, monitoring, and evaluating one's learning process~\cite{schraw2006promoting}.  Current works that define guidelines and design of technology to support self-reflection~\cite{bentvelzen2022revisiting} are built upon Schon's framework~\cite{eraut1995schon} that defines two types of reflection: (i) \textit{Reflection-in-action} which occurs during task execution, shaped by unforeseen outcomes; and (ii) \textit{Reflection-on-action} which refers to the reconstruction of actions based on our past memories and the conclusions drawn from them~\cite{slovak2017reflective}. 

Inspired by this framework, researchers have conceptualized and developed a wide spectrum of tools aimed at enhancing self-reflection, from those focusing on physical activity~\cite{botros2016go,cercos2016coupling} to conversational agents tailored for workplace reflection~\cite{kocielnik2018designing}.
These tools have been found to foster self-reflection by taking into account individuals' memories, social interactions and conversations~\cite{bentvelzen2022revisiting}. However, the context of these studies has been overwhelmingly the Global North; limited research exists regarding the availability and analyses of such intervention tools in the context of developing societies of the Global South, where the literacy on accessibility is minimal. Our study, which examines the unique experiences of BVI job-seekers in India, fills this knowledge gap.

\subsection{Interview Preparation and Feedback Tools for BVI Job-seekers}\label{RW2}
Constructive feedback before and after job interviews, regardless of the employment outcomes, is pivotal for enabling candidates to reflect upon their performance, pinpoint improvement areas, and refine their approach and skills~\cite{LinkedIn}. Similarly, self-assessment and ample practice during the preparation stage can elevate self-efficacy and reduce pre-interview anxiety~\cite{williams2008effects}. This has led to the conception and development of various tools designed to support job seekers, particularly those from vulnerable groups~\cite{hayes2015mobile,hendry2017homeless,hendry2017u,wheeler2018navigating,venkatraman2024you,prakash2023autodesc,prakash2024all,lee2022enabling}, in critically evaluating and receiving actionable feedback regarding their resumes~\cite{ResumeWorded}, skills~\cite{dillahunt2021skillsidentifier}, and professional self-presentation during interviews~\cite{hewitt2008impact}. For example, Dillahunt et al.~\cite{dillahunt2021skillsidentifier} created `SkillsIdentifier', a tool that detects skill gaps in underrepresented job seekers, and then aids in strengthening their resumes, thereby improving self-efficacy. The design of their tool was based on their prior study~\cite{dillahunt2016designing} which had explored the impact of such intervention technologies on meeting the needs of underserved job seekers. In another work, Hayes et al.~\cite{hayes2015mobile} utilized `VidCoach', a video modeling application, as an intervention for job-seeking students with Autism Spectrum Disorder (ASD), demonstrating that this tool could enhance interview performance by reducing anxiety and promoting coherent presentation of ideas. Recently, with the advent of advanced intelligent systems, several AI-based tools~\cite{AITools} have been developed, offering instant and comprehensive feedback to individuals seeking employment.

Recent CSCW works have also examined how BVI users perceive their roles and independence within an (in)accessible digital social context, offering valuable insights into their tools and strategies for addressing potential ableism in a high pressure interaction setting such as workplaces. Towards this, Saha et al.~\cite{saha2020understanding} studied blind audio professionals and observed how they relied on online communities as resources for learning, troubleshooting, and improving workflows. In another work, Lyu et al.~\cite{lyu2024because} studied how BVI users leveraged TikTok as a tool to share information and practice public speaking. Again, the context of all these aforementioned works is the Global North; the extent to which these tools are available or used in the developing countries like India, and the strategies employed by BVI individuals to prepare for interviews and to demonstrate competence in workplaces, are still open research questions.

\subsection {Job Seeking Practices of BVI Individuals}\label{RW3}
High unemployment rates coupled with the unique accommodation challenges posed by BVI people in workplaces, have widely inspired researchers in multiple domains to examine the barriers BVI individuals face while seeking employment~\cite{cmar2021job} and the strategies they typically employ to overcome them~\cite{grussenmeyer2017evaluating}. For example, Cmar et al.~\cite{cmar2021job} conducted an empirical study on the job-seeking practices of visually impaired youth. Their findings revealed that while many BVI individuals were interested in finding employment, they often did not actively engage in job search activities. However, this engagement improved significantly with targeted interventions and strong parental support. Additionally, the accessibility of the job application process and online portals plays a vital role in promoting active job search behaviors among disabled individuals. Lazar et al.~\cite{lazar2012investigating} investigated the accessibility and usability of online job application websites for blind users. Their study found that many online employment application processes were inaccessible, with only $28.1\%$ of application attempts being completed independently without any assistance. 

Grussenmeyer et al.~\cite{grussenmeyer2017evaluating} studied the accessibility aspects of interview process for BVI job seekers, noting significant challenges such as inadequate accommodations during interviews and inaccessible pre- and post-interview tests, which further increased the barriers faced by BVI individuals during the employment process. Furthermore, assistive technologies (ATs) play a significant role in enhancing the employability of visually impaired individuals by enabling them to perform computer based tasks that would otherwise be challenging. A study by Pal et al.~\cite{pal2012assistive} revealed how access to AT not only elevates confidence and independence among BVI job seekers in India, but also opens up a wider range of career opportunities. However, the high cost and limited availability of such technologies remain significant obstacles that need to be addressed to fully support BVI job seekers. 

While most prior studies have focused on the general job search barriers and practices of BVI job seekers in developed countries with greater access to ATs and intervention systems, none have specifically examined the antecedents of job search behaviors from a behavioral perspective, and that too in developing countries. In our study, we investigate the job search behaviors of BVI individuals from diverse economic and linguistic backgrounds in India, and uncover how they navigate the social dynamics of employment in an era characterized by high technological reliance.

\section{Methodology}
To understand the job-seeking behaviors and self-reflection strategies of BVI job seekers, we conducted an Institutional Review Board (IRB)-approved user study that involved semi-structured interviews. Details of the study are as follows.

\begin{table*}[t!]
\centering
{\def\arraystretch{1.1}
{\small
\begin{tabular}{ m{0.3cm} m{0.8cm} m{1.6cm} m{1.6cm} m{1.8cm} m{2.8cm} m{1.8cm} }
  \toprule
  \multirow{2}{*}{\textbf{ID}} & \multirow{2}{*}{\shortstack[l]{\textbf{Age/}\\ \textbf{Gender}}} & \multirow{2}{*}{\shortstack[l]{\textbf{Age of}\\ \textbf{Vision Loss}}}& \multirow{2}{*}{\shortstack[l]{\textbf{Education}\\ \textbf{Level}}} & \multirow{2}{*}{\shortstack[l]{\textbf{Employment}\\ \textbf{Status}}} & \multirow{2}{*}{\shortstack[l]{\textbf{No. Of}\\ \textbf{Job Applications}}} & \multirow{2}{*}{\shortstack[l]{\textbf{No. Of}\\ \textbf{Interviews}}}\\
  & & & & & &\\
  \midrule
  P1 & 32/M & Since birth & Masters & Employed & 150 to 170 & 4 to 5\\ \hline
  P2 & 23/F & Since birth & Undergrad & Employed & 30 to 40 & 6 to 7\\ \hline
  P3 & 22/F & Age 15 & Undergrad & Unemployed & 20 to 25 & None\\ \hline
  P4 & 28/M & Age 18 & Masters & Employed & 100 to 120 & 3\\ \hline
  P5 & 25/F & Don't know & Undergrad & Unemployed & 50 to 80 & 4 to 5\\ \hline
  P6 & 19/M & Don't know & High School & Unemployed & 10 to 20 & None\\ \hline
  P7 & 27/M & Age 7 & Undergrad & Unemployed & 200 to 250 & 10 to 15\\ \hline
  P8 & 23/M & Don't know & High School & Employed & 70 to 80 & 5 to 10\\ \hline
  P9 & 22/F & Don't know & High School & Employed & 50 & 1\\ \hline
  P10 & 19/M & Age 16 & High School & Unemployed & None & None\\ \hline
  P11 & 24/F & Since birth & Undergrad & Unemployed & 30 to 40 & None\\ \hline
  P12 & 35/M & Since birth & Masters & Employed & 500 to 600 & 20 to 25\\ \hline
  P13 & 28/M & Age 3 & Undergrad & Employed & 120 to 150 & 5 to 8\\ \hline
  P14 & 31/M & Don't know & Undergrad & Reemployed & 450 to 500 & 30 to 40\\ \hline
  P15 & 37/M & Don't know & Masters & Employed & 10 to 15 & 4\\ \hline
  P16 & 25/M & Since birth & High School & Unemployed & 200 to 300 & 7 to 8\\ \hline
  P17 & 29/M & Since birth & Masters & Reemployed & 40 to 50 & 3\\ \hline 
  P18 & 26/F & Age 10 & Undergrad & Unemployed & 50 to 60 & None\\ \hline 
  P19 & 26/F & Don't know & Undergrad & Unemployed & 30 to 40 & None\\ \hline 
  P20 & 22/F & Age 14 & High School & Unemployed & 5 to 10 & None\\ \hline 
  \bottomrule
\end{tabular}
}
}
\Description[Participant demographics]{The table contains information on participant ID, age/gender, age of vision loss, education level, employment status, Number of job applications, and number of interviews.}
\caption{Participant demographics. All information was self-reported by the participants.}
\label{table:evalparticipants}
\end{table*}

\subsection{Study Participants}
In this study, we aimed to engage both employed and unemployed participants who met specific eligibility criteria: (1) they must have visual impairment severe enough to need ATs such as screen readers or screen magnifiers to access digital platforms; (2) they had been actively pursuing employment opportunities in the digital industry for at least the previous six months; (3) they had a resume, either in digital or paper format; and (4) they had access to a device capable of connecting to the internet. We required prospective participants to respond to a set of screening questions to verify these criteria. For recruitment, we utilized both traditional offline methods and digital channels. 

We established partnerships with regional workforce development programs (training centers), following recommendations from earlier research~\cite{dillahunt2017uncovering}. Through these collaborations, we spread the word about our study via the programs' mailing lists and word-of-mouth. Additionally, we employed snowball sampling, where enrolled participants helped recruit additional participants. In our digital recruitment efforts, we targeted active Facebook groups that cater to BVI individuals, specifically those seeking employment. We prioritized groups with robust engagement, signified by a membership count exceeding $500$. This approach allowed us to tap into the vibrant online community and connect with potential participants who actively use social media as a resource for employment opportunities. 

From the pool of interested candidates, we selected $20$ participants who met all the inclusion criteria, including $8$ females and $12$ males. The age range of our participants was notably broad, extending from $19$ to $37$ years. The average age stood at $26.15$ years, with the median age at $25.5$ years, indicating that the distribution of ages has a few older participants that increase the average age. The standard deviation in age was $4.7$ years, highlighting the age diversity within our participant group. The participant demographics, detailed in Table~\ref{table:evalparticipants}, show a diverse representation of the BVI community.

\subsection{Study Design}
The interview questionnaire were meticulously crafted, taking into account established guidelines~\cite{passmore2002guidelines,batterton2017likert,kallio2016systematic,adams2015conducting} on job search process and insights from similar studies on job search behaviors in underrepresented populations~\cite{wanberg1999unemployed,akdur2022analysis,dillahunt2016designing,dillahunt2018designing}. The study commenced with an informed consent statement outlining the study's topic, a confidentiality pledge, permissions sought, an estimated time commitment, and an expression of gratitude. 
This was followed by a generic questionnaire on user demographics such as name, age, gender, employment status and history, visual condition, and education level. Next up was a semi-structured interview whose format allowed users to freely talk about their job seeking experiences and strategies.

The semi-structured interview aimed to investigate the job search behaviors of BVI individuals, how their experiences shaped their current job-seeking strategies, their self-reflective processes and their impact, and the effectiveness of job intervention tools in providing feedback and boosting self-efficacy. It commenced with inquiries regarding the participants' education and career aspirations. The next questions delved into their job-seeking and employment experiences, covering aspects such as interview experiences, challenges faced during the job-seeking process, self-reflection after rejection, attitudes towards the industry, and the assistance received during the job-seeking process. We then asked participants about any employment training tools they had used and their experiences with these tools. The questions also touched upon their perceptions of knowledge deficiencies and lack of skill sets that might hinder their employability (self-awareness). Analyzing responses across diverse employment demographics allowed us to gain insights into effective self-reflective strategies for successful employment prospects. Below are a few notable `seed' questions we asked the participants:

\begin{itemize}
    \item Can you describe your overall job search experience, including the number of jobs you have applied for, your interview experiences, and the support you have received from family, friends, and peers?
    \item Given the potential challenges of working in a digitally advanced environment with accessibility issues, how confident are you in your ability to adapt, and what strategies would you use to collaborate with sighted colleagues who might have limited knowledge about accessibility?
    \item Explain instances where you faced a setback or disappointment in your job search, and how this experience led you to reflect on and adjust your skills or approach to improve your chances of success?
    \item Have you used any digital tools or platforms to prepare for job interviews or practice skills-based questions? If so, how accessible and effective do you find these tools, and how do they compare to actual interview scenarios? 
\end{itemize}

We also posed follow-up and clarification inquiries for responses that caught our attention, including those that were novel, vague, evasive, or deviated from earlier answers. These questions were formulated in accordance with Hove and Anda's recommendations for conducting semi-structured interviews in empirical software engineering research~\cite{hove2005experiences}. To conclude the interview, we reiterated the research's purpose, outlined future work, and invited the interviewee to share any additional thoughts.

\subsection{Data Analysis}
The first author transcribed fourteen interviews conducted in English, while the second author transcribed the remaining six interviews conducted in Hindi and Kannada -- regional languages in India. The transcriptions resulted in a comprehensive book of $612$ single-spaced pages, with a maximum of $50$ lines per page. To analyze the collected qualitative data from the interview study, we utilized a hybrid process of inductive and deductive coding for thematic analysis~\cite{chung2017personal,boyd2016saywat,kameswaran2018we,kelly2017demanding,lanette2018much,mikalsen2018data,yarosh2016youthtube}. 
First, each researcher analyzed their respective transcripts through an inductive process using open coding. This involved a detailed line-by-line analysis of the interview transcripts to identify new themes and patterns that emerged from the data. The themes that emerged from this analysis were grounded in the interviewees' experiences; in-vivo codes that utilised the precise language of interviewees capturing the essence of their narratives. 
Following the inductive coding phase, we transitioned to the deductive phase. In this stage, we organized the emergent themes according to predefined theoretical frameworks. These frameworks were based on existing literature, specifically the Self-Regulation Theory (SR)~\cite{kanfer2001job}, which addresses topics such as job search self-efficacy and job search clarity, and the Theory of Planned Behavior (TPB)~\cite{van2009predicting,van2004job}, which encompasses topics such as job search attitude, subjective norms, and job search intention.
No pre-existing codes were used at the outset of the study; instead, codes were developed organically during the analysis process through constant comparison of the data and the application of labels to the text. After completing both phases of coding, we conducted a thorough review and comparison of the codes generated during the deductive and inductive stages. 
By combining inductive and deductive approaches, we were able to ensure that our analysis was both data-driven and theoretically informed. This hybrid approach allowed us to capture the richness of the interviewees' experiences while also situating our findings within a broader context.

\subsection{Positionality}
This research involved five authors, with the first, second, and fifth authors hailing from India, the fourth author from South Korea, and the third author from Sri Lanka. The first, second, third, and fifth authors identify as male, while the fourth author identifies as female. All researchers belong to ethnic minority groups in the United States of America (USA). All authors are sighted and work in the field of Human-Computer Interaction, with four authors specializing in accessibility research. 
The primary author led the data collection process, assisted by the secondary author, who is trilingual in English, Hindi, and Kannada. This language proficiency enabled participants who spoke these languages to comfortably share their experiences, allowing them to express their thoughts and feelings more effectively. 
The first two authors were involved in drawing insights from the semi-structured interviews, while the third and fourth authors contributed to refining the theoretical frameworks. All the study authors participated in interpreting the findings, discussing their implications, and outlining directions for future research.
Recognizing the importance of unbiased research, all authors took proactive measures to mitigate any assumptions or existing biases that could influence the study. The authors compiled lists of their preconceptions, openly discussing these to ensure they were aware of potential biases. This reflective practice was integrated throughout the research process, from data collection to analysis to interpretation.

\section{Context}

\subsection{Digital Transformation in Employment and Workspaces}\label{Context1}

The advent of Industry 4.0, a phase defined by the integration of cyber-physical systems, automation, and the Internet of Things (IoT), fueled by advancements in technologies such as artificial intelligence (AI), machine learning (ML), and advanced robotics, has transformed the job market into one marked by constant flux, evolving demands, and a relentless need for adaptation~\cite{li2022reskilling}. Innovations that once required decades to diffuse across industries are now being implemented within years or even months. The Indian Information and Communication Technology (ICT) job market which contributes to over 13\% to the country's GDP~\cite{ITA2024} has also adapted this change, with AI and ML fields creating nearly $69$ million new jobs and consequently a high demand for data scientists, cybersecurity experts, and cloud engineers~\cite{Deloitte2024}. However this accelerated shift has exacerbated an already significant and persistent skills gap~\cite{Deloitte2024,rathelot2023rethinking}. Reports from the World Economic Forum and McKinsey suggest that up to $50\%$ of workers will require re-skilling by 2025 due to the adoption of new technologies~\cite{Illanes2018}.

The COVID-19 pandemic further underscored the transformative impact of digitalization on the workforce. With the advent of COVID-induced lockdown, organizations worldwide had to transition to remote working models almost overnight, driving widespread adoption of digital collaboration tools such as Microsoft Teams~\cite{MicrosoftTeamsDocumentation}, Zoom~\cite{ZoomDocumentation}, and Slack~\cite{SlackDocumentation}. India's ICT sector played a crucial role in enabling this transition globally~\cite{deshpande2023long}. Indian IT companies like TCS, Infosys, and Wipro rapidly adapted to this shift, deploying massive remote workforces to support international clients~\cite{Reuters2024}. While skilled IT professionals in urban centers transitioned smoothly to remote work, workers in smaller towns and rural areas faced significant barriers due to inadequate access to technology and training~\cite{sharma2024covid}. This regional disparity was amplified by India's vast socio-economic divide, where only a fraction of the workforce had the digital literacy needed to adapt to the new demands of ICT-intensive roles. The transition disproportionately favored English-speaking professionals, further marginalizing workers from non-English-speaking and economically-disadvantaged backgrounds~\cite{sharma2024covid}.

The evolution in the industry has been complemented by advancements in ATs for BVI such as screen readers, text-to-speech software, and tactile devices, making career paths such as coding, data analysis, and digital marketing increasingly viable for BVI individuals~\cite{GeorgiaTech2023}. However, the rapid pace of industry evolution presents unique learning challenges for BVI individuals. For them, mastering new technologies involves not only understanding functionality but also navigating accessibility features~\cite{das2019doesn}. This dual learning process is both time-intensive and mentally taxing, often requiring specialized training resources that are frequently unavailable, particularly in developing countries~\cite{Pal2014}. In nations like India, a key hub for outsourcing work, providing BVI individuals with appropriate skill training and sustained upskilling support could significantly reduce unemployment~\cite{pal2012assistive}. With the right infrastructure and resources, this segment of the workforce could contribute meaningfully to high-skill industries, bridging gaps in inclusivity and employment.

\subsection{Accessibility in Job-Seeking and Employment Training in India}\label{Context2}

India has established itself as a global leader in ICT services with relatively lower labor costs, thus presenting an attractive option for companies seeking to reduce operational expenses without compromising on quality~\cite{kumar2022ict}. Programs like Digital India~\cite{DigitalIndia} and Skill India~\cite{SkillIndiaDigitalHub} have played a crucial role in promoting digital literacy and enhancing workforce skills, aligning them with global industry standards~\cite{kumar2022ict}. Efforts toward inclusivity have also gained traction, with organizations like the National Association for the Blind (NAB)~\cite{NABIndia} pioneering initiatives that provide specialized training in computer literacy and digital skills. These programs have enabled BVI individuals to access a broader range of employment opportunities. Similarly, initiatives like Vision-Aid’s Project Springboard~\cite{VisionAidProjectSpringboard} have expanded to multiple schools for the blind, focusing on empowering students through digital education and vocational training.

However, despite such skill development programs, the employment rate for BVI individuals remains alarmingly low. This disparity points to a disconnect between training initiatives and actual job placements, suggesting that existing programs may not adequately align with market demands or prepare BVI individuals for available roles. Additionally, the rapid pace of technological advancements presents challenges for training centers, which often struggle to keep curricula up to date~\cite{kharade2012learning}. The high cost and limited availability of ATs further restrict access to essential tools for digital education and employment~\cite{pal2012assistive}. Although India has enacted progressive legislation, such as the Rights of Persons with Disabilities Act (RPwD), 2016, which mandates a 5\% reservation in government jobs for persons with disabilities, these measures have not significantly improved employment rates~\cite{BlindWelfareSociety2023}. The Americans with Disabilities Act(ADA) in the U.S.~\cite{USDOL_ADA} and the Equality Act in the U.K.~\cite{EqualityAct2010} cover both public and private sectors comprehensively, enforcing accessibility and non-discrimination across all areas. While India's RPwD Act also applies to both sectors, the implementation and enforcement, particularly in the private sector, face challenges due to limited resources and awareness~\cite{USDOL_ADA,EqualityAct2010}. Key barriers include inadequacies in training programs, such as a focus on short-term courses that fail to impart employable skills, outdated curricula, inadequate infrastructure, and insufficiently trained instructors~\cite{TheWire2024}. These factors contribute to suboptimal training outcomes, leaving participants ill-equipped for the demands of modern industries. Furthermore, the lack of collaboration between training institutions and industries exacerbates the issue, as training programs often fail to align with current market requirements, perpetuating unemployment among BVI individuals~\cite{TheWire2024}. Hence, it is essential to understand and address the practical misalignment between BVI job seekers' perceptions of their skills and the actual demands of the industry. This can be achieved by fostering effective self-reflection on their abilities, facilitated through collective knowledge-sharing and constructive feedback mechanisms.

\section{Findings}

\begin{figure}[t!]
    \vspace{0.2cm}
    \includegraphics[width=\linewidth]{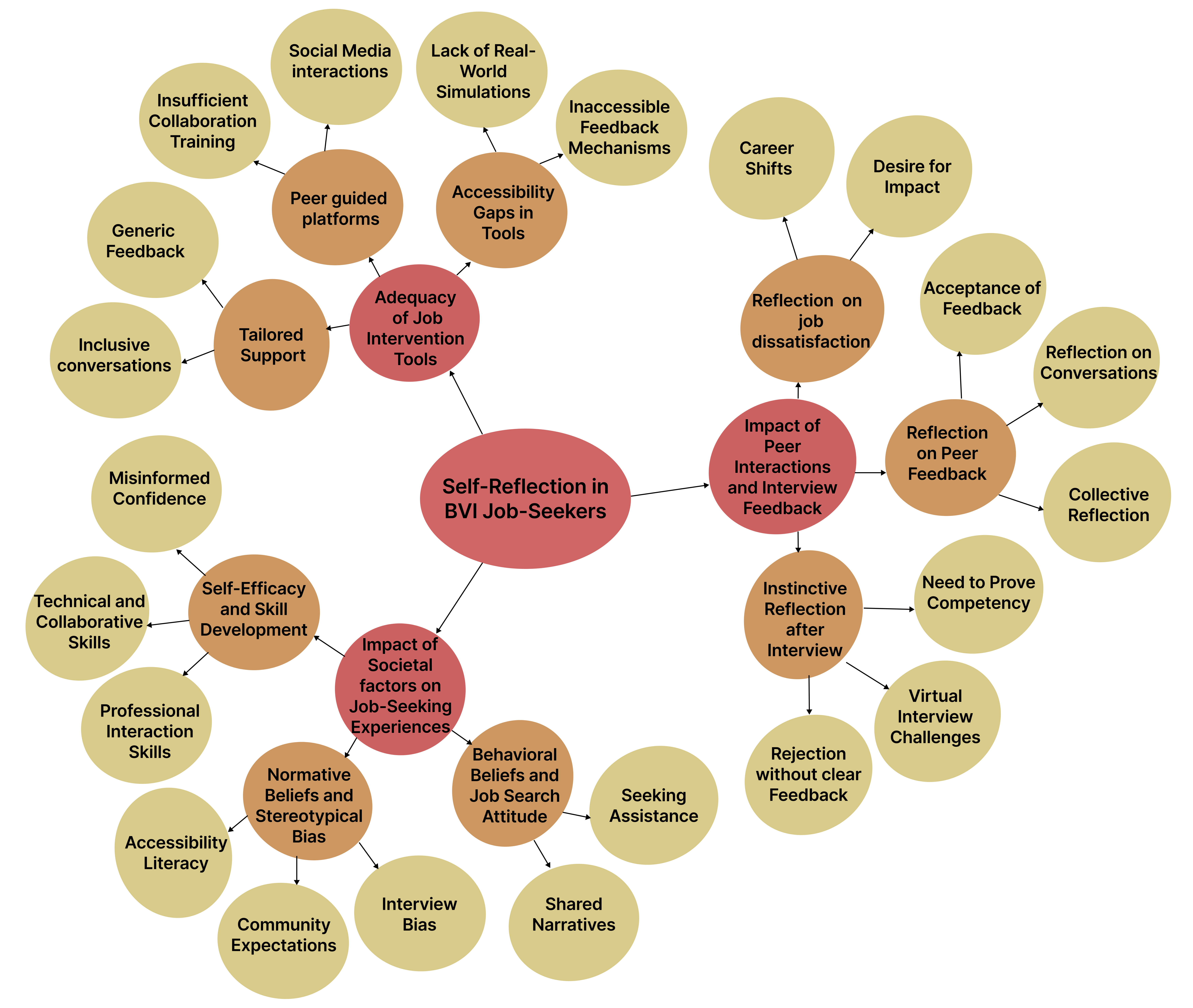}
    \caption{Key themes and sub-themes in self-reflection among BVI job seekers.}
    \label{fig:resultImage}
\end{figure}

Our analysis of interview data revealed three main themes: factors defining job search behaviors in BVI individuals, their self-reflection processes and subsequent decision-making, and the effectiveness of job intervention tools in providing constructive feedback. All themes and sub themes, illustrated in Figure \ref{fig:resultImage}, are detailed next.

\subsection{Impact of Societal factors on Job-Seeking Experiences of BVI Individuals (RQ1)}\label{antecedents} 

\vspace{5pt}
\noindent
\textbf{\textit{Self Efficacy and Skill Development.}} 
BVI individuals often develop higher levels of general self-efficacy due to the extensive mastery experiences acquired while overcoming challenges associated with their disability~\cite{brunes2021general}. This strong belief in their abilities is crucial in the job-search context, where individuals must effectively convey their competence to employers, demonstrating their capacity to perform work-related tasks and collaborate with diverse peers~\cite{dillahunt2020positive}. The participants in our study reinforced this notion, expressing increased confidence as they became more proficient with screen reader software and other digital platforms. This proficiency was often celebrated as a significant milestone by others in their society, where accessibility literacy was minimal. However, in the early stages of their careers, as they began the job-search process, participants realized that their confidence was often largely misinformed. They observed a significant disparity between the skills they collectively deemed sufficient and the practical demands of the industry. While proficiency in screen reader software allowed them to navigate daily tasks and communicate seamlessly with friends and family through digital devices, they lacked specialized job-related skills. Consequently, when faced with repeated rejections, they developed a negative attitude, often attributing their failures to societal discrimination and stereotypical misconceptions about their abilities. P7 described his experience: 

\begin{displayquote}  
    \textit{``I have been searching... for a couple of years now, attending about 10 to 15 interviews...I still haven't found a job. My skills match the positions... but the industry continues to undervalue my abilities because of my blindness. It's really disappointing to send out so many applications without... [receiving] any positive responses. They do not see the potential in people like me. I know I'm capable, and with the right opportunity and support, I can handle any work-related task proficiently.''}- P7
\end{displayquote}

Delving into specifics of mitigating this skills gap, participants, particularly those who were employed, reflected on how most educational institutions primarily focused on teaching BVI students to navigate the web and use platforms like Microsoft Office in isolation, providing little to no exposure to the collaborative features of these tools. For instance, P13 described how he had not learned to use these tools \textit{``in networking mode''} (referring to collaborative and cloud-based features). He explained, \textit{``Navigating these tools and working with sighted peers made me anxious. In [specialized] schools, we didn't get much chance to work with [sighted] people, so the whole experience was pretty new and intimidating for many of us.''} Other participants similarly described being educated primarily in specialized schools, where professional interactions with sighted individuals were minimal. While few others did work with their sighted peers who assisted them with their tasks, these interactions were often one-dimensional, in contrast to a work environment that requires mutual collaboration and collective ideation towards a common goal~\cite{das2021towards}.

Another unique challenge our participants faced was the need to simultaneously learn the functionalities of the software, its accessibility features, and the methods for collaborating synchronously with sighted peers using the tool. This learning process was continuous, as these software platforms were frequently updated and offered minimal explicit tutorials to support independent learning for BVI individuals. P4 explained \textit{``Each tool has multiple accessible features, but many of us are not aware of them, For example, Google Docs and Sheets have several built-in [accessibility] options that we are not aware of ... As new updates are released, new [accessibility] features are added, which makes it harder for us to [keep] track. [Training] centers often provide only basic training, which isn't enough ... There is a need for self-learning tools that can teach us how to use different software in an accessible way.''} Developing a comprehensive understanding of how to use these platforms synchronously with ability-diverse peers is essential for employment of BVI individuals in this sector~\cite{das2019doesn,das2021towards}.

In addition to mastering software-related skills, the current industry demands a range of soft skills, including effective communication, teamwork, interpersonal skills, conflict resolution, and time management, which are rigorously assessed during the interview process~\cite{nagarajan2008towards,valavosiki2019development}. While many of our participants were confident in their technical skills, they often struggled to establish clear and effective interactions with interviewers. P12, a computer instructor, said many of his students struggled to clear the job screening process due to communication difficulties. He said,

\begin{displayquote}  
    \textit{``We've had numerous recruiters come to hire my students, but most... could not clear the interview due to lack of communication skills. In technical interviews, blind candidates often receive... a bit more leeway, with lower cutoffs compared to their sighted peers. But when it comes to communication skills, it's a different story entirely... Companies expect good proficiency in English, and that's where many struggle. Effective communication is crucial, it's something we really need to focus on. ''}- P12
\end{displayquote}

Referring to the support available for developing these skills, participants highlighted that while most computer training centers provided basic digital literacy, they rarely focused on teaching collaborative and team communication skills. This gap was attributed to two key factors: many instructors lacked direct exposure to the ICT industry, and the institutions had minimal interaction with professional recruiters. Online tutoring platforms, while potentially accessible, lacked specialized courses tailored for BVI job seekers and rarely addressed the intricacies of sighted-BVI communication. Instead, they typically provided generic, one-size-fits-all instruction for their audience.

The aforementioned scarcity in collaboration and communication-specific skill development, combined with repeated rejections, adversely affected the self-efficacy of our participants during their early careers, fostering a sense of alienation in the ICT industry. Consequently, many participants sought job opportunities in government sectors, which offered long-term financial stability, enabling them to support themselves and their families.

\vspace{5pt}
\noindent
\textbf{\textit{Normative Beliefs.}} The shared expectations and collective knowledge gained through social interactions with friends, family, and other significant individuals--often referred to as an individual's `normative beliefs'~\cite{ajzen2006behavioral}--can define the effectiveness with which they engage in the job search process~\cite{dillahunt2020positive}. Reflecting on their early years, participants emphasized how the collective knowledge within their community about digital accessibility and assistive technologies shaped their initial career options. These choices were often influenced not by personal interests but by what was traditionally regarded as \textit{``possible''} and \textit{``respected''} by their family and peers. While our participants' families were supportive of their education, they often lacked the knowledge or resources to guide them in accessible digital navigation, thereby limiting their career possibilities. P16 described,

\begin{displayquote}  
    \textit{``I grew up in a village where they[parents] taught me at home. They believed that government roles, like office jobs or teaching, were not only accessible but also respected and prestigious career options for someone blind like me... It wasn't until I moved to Bengaluru after high school to attend a specialized school that I was introduced to screen readers and various online platforms. Although this has opened up many new opportunities for me, gaining access to these tools earlier... would have helped in my school and helped in choosing my career path better.''}- P16
\end{displayquote}

While participants lacked early exposure to technology, they reported no trouble adapting to the social dynamics with sighted peers during their education. This is crucial, as schools create collaborative learning environments that encourage students to articulate their thoughts, challenge differing perspectives, and co-construct knowledge, leading to deeper understanding and retention of material while also fostering social competencies among students~\cite{tenenbaum2020effective,hank2024peers}. Among our participants, $14$ attended specialized schools for BVI individuals, while $6$ were educated in regular schools alongside sighted students. Participants highlighted how the inclusivity and peer support they received, even when sighted peers did not fully grasp the specifics of BVI accessibility, enabled them to pursue opportunities on par with their sighted classmates. Additionally, competing with sighted individuals where they often found accessible alternatives to do the same tasks helped them look beyond their disability, recognizing that they could aspire to the same goals.

Participants who attended specialized schools reported having greater access to peer mentorship, where senior students and BVI teachers often guided them in navigating accessibility, education, and later, the job search process, drawing from their own prior experiences. These mentors frequently ``shared tips'' about using screen readers and other assistive technologies, providing an informal learning avenue and fostering a comfortable environment for collective skill development.

While BVI participants graduated with high levels of self-efficacy regarding their skills and their ability to project a positive self-presentation in interviews, they often failed to pragmatically reflect on the expectations of a possibly biased interviewer. Participants who disclosed their visual condition during the application process often described carrying the \textit{``burden''} of convincing recruiters of their ability to fulfill job responsibilities independently and without significant assistance.

When interviewers focused excessively on the abilities of BVI participants and occasionally expressed surprise at their qualifications, albeit inadvertently, it was perceived negatively by the participants. The need to constantly persuade recruiters during every interview made the process feel more daunting, leaving participants with a sense of diminished control over their job search as a whole. Additionally, many recruiters expressed a lack of confidence in their ability to integrate employees with visual impairments, fostering a sense of inevitability and discouragement among the participants. P9 shared, \textit{``In one interview, the recruiter said they'd have to spend extra on licensed screen reader software. They also mentioned they might not have the proper training facilities to teach me new software platforms they might use. Hearing this kind of feedback was really discouraging, making many of us [BVI] lose hope and look for jobs outside of computer-dominated fields.''}

While we acknowledge that fully mitigating sterotypical bias regarding capabilities on BVI individuals through broader education on inclusivity and accessibility will take time~\cite{whysall2018cognitive}, it is crucial to train BVI job seekers to effectively navigate these situations during interviews. Additionally, guiding them to reflect on their skills independent of the outcomes or negative interactions is essential to prevent a decline in their self-efficacy. We explore this in greater detail in the Discussion section.

\vspace{5pt}
\noindent
\textbf{\textit{Behavioral Beliefs and Job Search Attitude.}} 
The effectiveness and intensity with which individuals pursue employment are broadly influenced by attitudinal factors, including their thoughts, judgments, and feelings toward employment~\cite{ajzen2006behavioral,kanfer2001job}. Shaped by prior interview interactions, participants frequently perceived a negative impact of their visual condition on the process, anticipating that their disability would often overshadow their qualifications. This expectation of a negative interview experience affected their job search intensity. Some participants reduced the number of applications they submitted or targeted only organizations known for being accommodating. However, those who had a strong foundation of continuous interaction with sighted peers during their school days expressed a shared belief in working collaboratively with sighted colleagues and contributing productively towards a common task. When it came to seeking assistance, participants trusted that most people were willing to help when asked, and this belief encouraged them to proactively seek guidance or information when needed.

However, $5$ participants indicated a habitual reliance on sighted peers and a tendency to withdraw from tasks if they perceived insufficient assistance in completing them. For instance, P2 described that she would \textit{``talk to the manager and step back''} from a task or even request reassignment in cases where she anticipated a lack of proficiency, which would require her to seek constant assistance from others.

When discussing their job preferences, participants emphasized that their choices were often shaped by the narratives and collective experiences of the BVI community, which steered them toward employment sectors perceived as more welcoming and attainable. For example, participants mentioned jobs in call centers and Non-governmental organization (NGOs), where positive feedback from senior peers who had secured positions fostered trust that these employers would be more understanding of their needs and willing to provide necessary accommodations. Conversely, narratives about the IT sector, which depicted it as unwilling to adapt and accommodate BVI job seekers, contributed to a negative outlook. This perception led participants to approach interviews in the IT industry with lower expectations and diminished confidence, believing that the effort required to enter the sector might not yield proportional rewards.

Interestingly, the participants stated that they also developed a pragmatic understanding through interactions with experienced peers, recognizing the accommodations and additional investments in ATs that employers would need to make to hire them. Participants often reflected on the likelihood of these investments being made by different companies and reported proactively discussing these considerations with interviewers during the recruitment process.

\subsection{Impact of Peer Interactions and Interview Feedback on Self-Reflection and Career Decision-Making (RQ2)}\label{result2}

Participants recounted various instances where interactions with peers and recruiters during the interview process instinctively prompted them to reflect on different aspects of their job search. They emphasized how this reflection often fostered a more proactive and strategically informed job-search approach while helping them navigate adverse outcomes. 

\vspace{5pt}
\noindent
\textbf{\textit{Reflection on Peer Feedback and Conversations.}} BVI participants who were successfully employed reported establishing networks with both sighted and BVI individuals. These networks provided a diverse set of constructive inputs, not only on the job search process but also on understanding the social dynamics of interviews and workplace interactions. P19 advocated for an iterative approach of \textit{``receive,'' ``reflect''} and \textit{``act,''} emphasizing the importance of developing an understanding of how sighted peers with no prior experience of interacting with BVI individuals or knowledge of accessibility might perceive and respond to them. P9 described, 

\begin{displayquote}
    \textit{``It is important to really tune into chats with sighted people to get our communication right. Here in India, a lot of people don't know how to interact with us. It's not like they are discriminating or anything ... it's mostly just not knowing better. This makes them either too cautious ... almost like they would [rather] not talk much. And in interviews, where you really need to talk with the recruiter, [smooth] communication becomes even more important.''}- P19
\end{displayquote}

Participants described how they often recalled and reflected on conversations with their sighted peers to identify any contextual misalignments. These reflections were frequently shared and discussed with their BVI peers, allowing them to exchange experiences and thoughts on how to navigate and feel more comfortable in such scenarios, particularly during formal interactions with higher-ups or when communicating with someone new via videoconferencing. This process highlights how reflection can be a collective endeavor, with social interactions serving as a significant catalyst for learning and adaptation.

Another key factor for effective workplace outcomes is not just constructive interactions but also the manner in which feedback is received or accepted. It is important to avoid negative emotions, such as questioning the accuracy of the feedback or the qualifications of the provider, as these reactions could lead to unnecessary disagreements and hinder productivity~\cite{gnepp2020future}. Our participants reported that they initially felt a sense of inferiority when receiving negative feedback, making it harder for them to accept it without feeling judged or undermined. They often perceived such criticisms as reflective of stereotypical biases held by their sighted peers rather than as objective assessments of their abilities. P8 described,

\begin{displayquote}  
    \textit{``Many of my students get really annoyed when they are criticized, especially when it comes from a sighted person. They think that the sighted people do not have the same understanding or experience with blindness, so the feedback can feel unfair...This makes the whole [learning] process more frustrating for them..I tell them that it is important to communicate their ideas[perspectives] and teach their sighted peers, but it is tough because they feel like they are constantly having to prove themselves.''}- P8
\end{displayquote}

This sentiment among BVI individuals was rooted in prior experiences with employer prejudice and a persistent feeling of being misunderstood and unfairly judged. These experiences often made constructive feedback appear unfair or uninformed~\cite{heydarian2022brief}. Participants preferred to trust their \textit{``gut feeling''} and \textit{``make their own decisions''}, over advice from friends and even family at times. However P12 described that this preconceived notion hampered his students' ability to logically comprehend and derive reflective insights from feedback. Consequently, self-reflection had become a more complex and challenging process for them.

One more key factor influencing how BVI participants engaged in self-reflection was the nature of the environment and interactions in which they received feedback. Participants indicated that they learned effectively in informal settings with peers, with whom they spent most of their time. These interactions often included discussions on various topics such as ATs, employment opportunities, and updates on day-to-day activities. However, this dynamic changed as they transitioned to more professional settings, where interactions involved individuals with minimal understanding of their experiences, leading to feelings of disconnection. P4 described,

\begin{displayquote}  
    \textit{``I traveled from Delhi to Bengaluru for computer training.... [spent over] a year at the training center. [While] I learned a lot from my peers, the job market was a whole different story. [At the training centers], we did not face real job scenarios, which made it hard to understand workplace expectations. It was only when we faced actual job situations that we realized what more we needed to learn.''}- P4
\end{displayquote}

The participants also noted that while their time at training centers equipped them with the basic skills needed to handle routine computer tasks, the lack of exposure to diverse team-activity scenarios inadvertently perpetuated the existing skills gap. This gap was particularly evident in their ability to interact effectively, communicate ideas, and receive constructive feedback from others in a professional setting.

\vspace{5pt}
\noindent
\textbf{\textit{Instinctive Reflection after Interviews and Job Rejections.}}
Interviews are pivotal moments that prompt candidates to introspect on their skills, assess their ability to convey thoughts clearly, articulate ideas effectively, build rapport with the interviewer, and demonstrate enthusiasm for the role and the company~\cite{ResumeheadSelfReflection}. Reflecting on their interview experiences, participants described being highly conscious of how they expressed themselves, often second-guessing whether they were communicating their answers clearly to the interviewer. They paid close attention to cues that could help them adapt their responses in real time. P4 described,

\begin{displayquote}  
    \textit{``Confidence is key for us during interviews. Recruiters often doubt if we can handle job responsibilities without help. So, during the conversation, I [am] always paying attention, wondering if I'm proving my abilities...It is a constant process of checking myself and adjusting to make sure I can show my competence.''} - P4
\end{displayquote}

This process of second-guessing in a high-pressure interaction scenario made the experience even more daunting, particularly in virtual settings where additional factors had to be considered. This was explained by P2,

\begin{displayquote}  
    \textit{``Zoom meetings are even tougher for me. I am always worried if my body posture is right...if my face is visible...if my internet [connection] is stable...if I am looking at the camera correctly... I am also anxious about [whether] my screen reader is picking up everything accurately.'' - P2}
\end{displayquote}

The effectiveness with which participants engaged in self-reflection after job interviews was significantly influenced by the quality of the interaction and the feedback received, regardless of the outcome. Participants often reported being rejected with vague explanations and without clear feedback, leaving them with no direction for self-assessment or skill development. In the absence of justifiable feedback, many participants perceived the rejection as stemming from stereotypical biases or an unwillingness to hire them.

After facing multiple rejections, participants often decided to explore alternative career options, as they were unable to identify the specific skill gaps that hindered their employment and had no clear direction for improvement. Many participants reported reaching out to employed friends for personalized feedback on their job search approach. They frequently recalled and discussed their interview experiences to reflect on how effective their interactions were and how their communication might have come across to interviewers, whom they believed often lacked a full understanding of BVI perspectives and capabilities.

Participants also practiced mock interviews with friends, both sighted and BVI, which provided broader perspectives. However, this method had limitations. Access to such support was not always readily available, as many participants expressed reluctance to \textit{``continuously disturb''} their friends. Moreover, practicing in informal settings did not always simulate the high-pressure, professional environment of actual interviews. Additionally, in many instances participants reported that their peers hesitated to provide critical feedback, further limiting the effectiveness of this approach.

\vspace{5pt}
\noindent
\textbf{\textit{Reflection on Job Dissatisfaction and Pursuit of Self-Actualization.}} Drawing inspiration from Maslow's hierarchy of needs~\cite{bacslevent2013preferences, maslow1987maslow}, we explored how job dissatisfaction and the desire to find meaningful work could have triggered self-reflection among BVI job seekers. Participants expressed a strong collective desire not only to achieve financial independence but also to uplift their community by creating employment opportunities for fellow BVI individuals facing similar challenges. Reflecting on their own journeys—from remote backgrounds with limited access to computer education and ATs, battling societal stereotypes, and overcoming job market hurdles—they felt a shared responsibility to support others in their community. P17 described,

\begin{displayquote}  
    \textit{``Sir, do you know KK Mane? [Referring to Krishnakant Mane, India's first visually challenged IT professional and technology entrepreneur~\cite{peoplematters}] I want to be like him one day. I've been working in IT as an accessibility tester for a year now but I plan to quit and start my own organization that can provide jobs to many people from my village and community like me.''} - P17
\end{displayquote}

This sentiment among participants underscores how success stories, popularly shared within the BVI community through word-of-mouth, provided tangible examples of what was achievable. Participants expressed that by establishing successful careers, they could advocate for more inclusive hiring practices and showcase the potential of BVI individuals in various professional fields. They highlighted how reflection was often a social process, wherein hearing about one person's success led them to reflect positively and develop a collective sense of confidence and aspiration. This communal reflection fostered a shared belief in the potential for success, encouraging more BVI job seekers to pursue their career goals despite the challenges they faced. While this reflective process had a positive impact, some participants noted that it needs to be well-informed and regulated to avoid unrealistic expectations. P4 said,

\begin{displayquote}  
    \textit{``Many of my students want to start their own business or often reject job offers in the digital sector, waiting for government jobs, which they believe are more noble. It's crucial that they understand the realities of both options and make informed choices. ''} - P4
\end{displayquote}

Employed participants also expressed uncertainty and dissatisfaction with their current employment, as they were often hired on a contractual basis, which naturally led to a common fear of instability. When these concerns were shared with others seeking jobs, it prompted reflections on the substantial effort required to secure these positions, only to face the reality of no guarantee of long-term employment. Participants reported that this negative reflection often diminished their motivation to continue the job search.

\subsection{Adequacy of Current Employment Intervention Tools as Facilitators of Self-Reflection (RQ3)}

\vspace{5pt}
\noindent
\textbf{\textit{Accessibility Barriers in Employment Intervention tools.}}\label{access} 
BVI individuals rely on screen reader AT such as JAWS~\cite{jaws} and NVDA~\cite{nvaccess}, which read aloud on-screen content and provide one-dimensional navigational support via keyboard shortcuts. The compatibility of employment intervention tools with these ATs is therefore crucial. Without such compatibility, these tools can introduce additional cognitive burdens, potentially hindering the learning process altogether. Participants listed several tools they explored based on suggestions from their sighted peers. These included tools for resume refinement, AI-based interview feedback, and traditional interview record-review platforms. For instance, P8 and P4 said,

\begin{displayquote}  
    \textit{``I read about an AI-based interview prep tool on Google and decided to try it because it offered voice-based input, which made me think it would be accessible. The big problem was that my screen reader picked up both its own voice output and mine, making the AI's responses confusing. I tried lowering the system volume, but that didn't help. The whole thing ended up being really frustrating and completely unusable.''} - P8 
\end{displayquote}

\begin{displayquote}  
    \textit{``I tried using a practice interview platform that records mock interviews and lets you share them with others. I thought it would be great to record my responses and get feedback from my friends. But during the practice, my screen reader couldn't even navigate to the questions. ''} - P4 
\end{displayquote}

In a society with minimal accessibility literacy, participants had hoped that these tools would help them reflect on their skills and provide unlimited access to unbiased feedback. However, they often faced significant challenges overcoming the accessibility limitations of these tools. In cases where the platforms were accessible, BVI participants reported benefiting greatly. For instance, P5 mentioned,

\begin{displayquote}  
    \textit{``For general interview practice, I found Google Warmup really helpful I loved how I could easily navigate the entire platform with simple shortcuts, making everything so accessible. The insights on my talking points and the terms I frequently used were really useful. I would take notes, review them to see what needed improvement, and then try to implement those changes in my next interview It would be fantastic if it also included support for blind-specific job roles like accessibility testing and content writing.'' - P5} 
\end{displayquote}

Participants reported perceiving the feedback from such accessible tools as \textit{``unbiased''} and \textit{``fair,''} with minimal concern about being judged based on their disability. They interpreted the feedback more objectively, and critically reflected on their skills, including their communication and language. 

Participants also suggested several usability enhancements that could have significantly improved their experience with interview practice tools. For instance, P5 suggested the inclusion of a \textit{``repeat button''} for questions within the tool. This feature would simulate a realistic interaction scenario where users could ask the virtual interviewer to repeat a question they did not understand, mirroring real-life conversations. Similarly, P11 requested \textit{``training sessions or demo videos''} to guide them through using these tools and during their practice interviews. These resources would provide step-by-step assistance, helping users navigate the platform more effectively.

\vspace{5pt}
\noindent
\textbf{\textit{Collaborative Reflection through Peer-Supported Platforms.}} 
Other key factors influencing how participants reflected on their skills and job-search process were the information and feedback they received from friends and others in their community via social media platforms. These platforms enabled participants to extend beyond their less-informed immediate social circles and access a global network, where they could engage with professional groups and discussions while also receiving emotional support critical for building positive self-efficacy in their job search. Social media also informed them about advancements in ATs and opportunities within their own society, which were often under-publicized without these platforms. For instance, P13 described,

\begin{displayquote}  
    \textit{``I'm on my phone a lot, reaching out to people in Facebook groups and on WhatsApp, asking around about job openings or what kind of roles I should go for. It's how I landed at my current job. Hearing from others, getting their take on things, it makes the whole job prep feel more hopeful, less of a solo ride.'' - P13} 
\end{displayquote}

Discussions on these platforms not only offered participants insights into plausible job opportunities across the globe but also provided venues to relate their own experiences and collectively reflect on potential changes they could make to improve their job search approach. P6 noted,

\begin{displayquote}  
    \textit{``I really enjoy reading Reddit discussions. It's encouraging to see stories from other normal people who've succeeded in tough fields like IT. It makes me feel less like I'm fighting a losing battle against a world that doesn't understand me. I also read a lot about how different people, sighted or BVI, interact in various settings. When their stories sound like mine, it helps me feel more confident and gives me ideas on how to handle my own social situations better.''} - P6 
\end{displayquote}

\vspace{5pt}
\noindent
\textbf{\textit{Lack of Tailored Support for Effective Self-reflection.}} While participants drew inspiration for self-assessment from various sources, they lacked personalized feedback support crucial for effective self-reflection~\cite{dillahunt2020positive}. In the absence of this tailored information, participants reported often struggling to fit the generalized recommendations to their unique experiences, and even when they did, they were not always effective in meeting their practical employment needs. P1 explained,

\begin{displayquote}  
    \textit{``Friends keep saying, 'Pick up new skills, be confident, stay calm in interviews.' They mean well, but it's not that simple for me. At interviews, the pressure's intense. I feel underestimated. And that anxiety? It wipes the slate clean of all those tips. Even coaching centers they do cover tech skills, but there's little attention to what each of us really needs, or how we are doing individually. They don't see it.'' - P1} 
\end{displayquote}

Participants expressed a similar sentiment towards the feedback they received from the employment preparation tools they had tried using in the past. They mentioned that while these tools provided insightful suggestions, they did not take into account the complexities and requirements introduced by their visual disability or their unique educational backgrounds. Participants often found a significant mismatch between the questions they trained on using these tools and the actual ones posed by recruiters during interviews. P9 explained,

\begin{displayquote}  
    \textit{``I've been using that interview prep tool for a bit, really focusing on the feedback it gave. But, you know it didn't have anything for the accessibility testing job I was after. Then in the interview, totally different scene — they're asking all these critical questions about using accessibility features on web tools, especially where my screen reader might just give up. None of that was in the prep tool, caught me completely off guard, felt like I prepped for the wrong exam or something.'' - P9} 
\end{displayquote}

Participants also highlighted challenges in interacting effectively with these tools. Growing up in a linguistically diverse Indian society where English is not the first language, many participants had only a basic grasp of English and often struggled to understand the questions posed by the tools. Furthermore, a few participants noted that certain words used in the suggestions provided by these tools would sometimes be uncomfortable or confusing. These instances often made them feel that the tools were not designed with the needs of people with disabilities in mind. P3 mentioned,

\begin{displayquote}  
    \textit{``You know, I got some advice once that said I should `maintain eye contact.' That was a bit frustrating...Another time, it suggested... [I] `watch my body language.' Comments like these? They're just confusing, not helpful at all.'' - P3} 
\end{displayquote}

These comments highlight that merely making the tools accessible is insufficient to meet the unique requirements of BVI job seekers. The entire interaction pattern and feedback system must be designed with consideration for their real-life experiences, demographic and cultural backgrounds, and current skill-sets.

\begin{table}[t!]
\centering
\renewcommand{\arraystretch}{1.5}
\setlength{\tabcolsep}{8pt}
\begin{tabular}{@{}m{3.4cm}m{5.3cm}m{5.3cm}@{}}
\toprule
\textbf{Subject} & \textbf{\hspace{1.5cm}Findings} & \textbf{\hspace{1.5cm}Design Strategies} \\
\midrule

Promote Social Networking for Digital Literacy &
\begin{itemize}
  \item BVI job seekers face a skills gap due to limited educational opportunities, lack of societal support, and low collective awareness of accessible technologies.
  \item Repeated, unexplained rejections diminish their sense of control and self-efficacy.
  \item Demonstrating independent digital proficiency during interviews adds cognitive strain.
\end{itemize} &
\begin{itemize}
  \item Encourage early engagement with global networks to improve awareness of accessible technologies and job prospects.
  \item Offer training in soft skills to help BVI job seekers confidently express their needs and competencies in professional interactions.
  \item Use success stories to inspire positive self-reflection and support ongoing skill development despite adverse outcomes.
\end{itemize} \\
\midrule

Seek Personalized Feedback for Skill Development&
\begin{itemize}
  \item Peer-feedback is often too generic and disconnected from their lived real-life job search and interview interaction experiences.
  \item BVI individuals often receive minimal feedback during interviews and are rejected with vague explanations, leaving them without clear guidance for self-reflection and skill development. 
  \item Early BVI job-seekers struggle to objectively reflect on critical feedback.   
\end{itemize} &
\begin{itemize}
    \item Promote group discussions and knowledge sharing on social media platforms to facilitate collaborative reflection on constructive feedback.
    \item Simulate realistic interview scenarios for training using AI-assisted technologies to provide actionable feedback on professional interaction skills.
    \item Consider the user's background, job-search history, and experiences to personalize feedback effectively.
\end{itemize} \\
\midrule

Accessibility of Intervention Tools &
\begin{itemize}
  \item Most employment intervention tools are not fully compatible with assistive technologies and present significant usability challenges.
  \item Diverse socioeconomic and linguistic backgrounds influence the accessibility of intervention tools.
  \item Tools often lack tailored support and impose significant cognitive strain on users due to accessibility issues.
\end{itemize} &
\begin{itemize}
  \item Redesign tools to enable seamless interaction with assistive technologies like screen readers.
  \item Design multilingual collaborative learning platforms to support diverse users.
  \item Integrate AI-driven systems to offer personalized language learning and communication skill improvement.
\end{itemize} \\
\bottomrule

\end{tabular}
\caption{Summary of design strategies correlated with findings and subjects.}
\label{tab:design_guidelines_summary}
\end{table}

\section{Discussion}
Our exploration of how BVI job seekers self-reflect in response to adversities in the job search process and the feedback they receive, within a society with limited understanding of their experiences and digital navigation capabilities, provides valuable insights for designing intelligent intervention tools and design strategies (summarized in Table \ref{tab:design_guidelines_summary}). These tools can guide them towards a more productive and regulated job search process. Some of the notable insights are discussed next.

\subsection{Peer-guided Self-reflection for Effective Skill Development}
BVI individuals, who remain a marginalized group in Global South countries such as India, face structural challenges including limited access to education, discrimination, socioeconomic constraints, and a lack of collective understanding of accessible technologies (refer Sections \ref{RW3}, \ref{Context2}, and \ref{antecedents}). Therefore, the design of interventions should consider their unique subjective experiences and align with their lived realities~\cite{hojdal2020contemporary}. To this end under RQ1, we extended prior literature that broadly explored the common challenges BVI individuals in job searches. We delved into the nuances of how BVI job-seekers in India  reflected on their experiences and interactions, and how biased societal beliefs shaped their perspectives regarding employment in the digital sector. We found that, due to a lack of general literacy of accessibility and limited exposure to professional interactions and collaborative settings, BVI individuals often struggle to effectively reflect on their skill gaps. As a result, they face challenges in meeting the rigorous demands of the industry, often attributing early career rejections entirely to societal prejudice, which diminishes their self-efficacy and perceived control over the job search process (Section~\ref{antecedents}). Additionally, the need to \textit{``prove''} their competence during the interview process imposed a significant cognitive burden, diminishing their ability to convey their thoughts clearly—a scenario for which they have been entirely unprepared.

While prior research has broadly advocated for measures to overcome systemic inequities in the employment of underrepresented job seekers~\cite{dillahunt2020positive}, it is equally crucial to promote individual adaptation for successful employment outcomes~\cite{hojdal2020contemporary,dillahunt2017uncovering,dillahunt2020positive}. A key aspect of this is fostering positive self-efficacy through effective and structured self-reflection, which is deeply embedded in social interactions that significantly shape and inform the reflective process. Structuring this self-reflection through well-informed facilitators, whose feedback encourages deeper analysis of BVI job seekers' prior experiences and beliefs, is essential~\cite{berry2021supporting}. However, the quality of self-reflection among these individuals is often limited by the knowledge and experiences of their immediate peers. In a society where a pervasive lack of collective understanding about the potential of BVI individuals in digital workspaces exists, more informed and supportive perspectives remain scarce. Here, we draw parallels to the social concept of the \textit{``looking-glass self,''} which posits that individuals form their self-identity based on their perception of how others view them~\cite{cooley1902looking}. For BVI job-seekers, feedback received over social networks serve as \textit{``mirrors''} helping shaping their professional identity and sense of capability. These social networks can potentially act as critical data sources~\cite{schwanda2012see} that offer more accurate, empathetic, and expertise-driven feedback that guide self-reflection in a positive direction. These networks can therefore help the users see themselves beyond the limited lens of immediate peers, who may share similar uncertainties or misconceptions. 

In the second part of our study (regarding RQ2), we explored the quality of feedback practically available to our BVI participants. We found that such feedback is often not readily accessible, and when available, it tends to be too generic and disconnected from their real unique experiences. Additionally, BVI individuals, often from economically and linguistically diverse backgrounds, may develop good communication skills in informal settings but experience significant anxiety during interviews and struggle to objectively receive feedback from interviewers and peers. In the absence of meaningful feedback, they internalize negative experiences and develop inaccurate self-assessments, either overestimating or underestimating their abilities (Sections \ref{antecedents} and \ref{result2}). Establishing participation in global peer networks could enable BVI individuals to seek tailored feedback on their skills and approach. Hence, we advocate for educators to teach BVI students how to comfortably engage with and receive constructive feedback while filtering out negative elements on these social platforms. Early self-reflection through such networks can help them make more informed career choices and approach professional interactions with greater confidence.

\subsection{Rethinking the Design of Employment Intervention Tools}
As established in prior research (Section \ref{RW2}), intervention tools that promote skill development by incorporating positive and constructive feedback are essential for helping marginalized job seekers, such as BVI individuals, maintain positive self-efficacy throughout the employment process despite the negative experiences and social prejudice they face. These tools are crucial for enabling them to make more informed career decisions, build professional collaborative skills and positive self-representation in interviews particularly in the absence of adequate societal support. Under RQ3, we examined the adequacy of existing tools in addressing the unique challenges faced by BVI job seekers in Global South countries (Sections \ref{antecedents}, \ref{result2}, \ref{Context1}, \ref{Context2}), particularly in the context of rapid technological evolution (Section \ref{Context1}). In addition to lacking accessibility support (Section \ref{access}), these tools fail to simulate realistic interview experiences for BVI individuals, lack tailored support, and are not designed to facilitate BVI-inclusive and professional interactions. This necessitates designing beyond a one-size-fits-all approach, creating tools that are usable, constructive, and tailored to the realistic experiences of BVI individuals in under-informed communities. Building on prior research and our findings, which offer a deeper understanding of the unique challenges faced by BVI individuals, we propose design suggestions to be incorporated in the development of intervention tools tailored to their needs.

Firstly, traditional tools such as job search platforms, resume builders, and interview preparation tools, which heavily rely on GUI-based visual feedback, should be redesigned to enable seamless interaction via assistive technologies (ATs) like screen readers and screen magnifiers~\cite{prakash2024improving}. Our research highlights that accessibility issues in these tools can impose significant cognitive strain on users, which can undermine their confidence in technology—contrary to the intended purpose of promoting self-efficacy among BVI individuals for employment in digitally collaborative workplaces. Moreover, making technology accessible does not inherently make it usable for BVI individuals~\cite{das2019doesn,prakash2024all,prakash2024towards}. These tools should adopt a simplistic design that minimizes the learning effort required to use them~\cite{prakash2024all}, ensuring that users do not need to seek external assistance or develop workarounds to address accessibility issues. The design of such systems must follow a user-centric approach, involving rigorous testing for both usability and the cognitive burden they induce during interaction. Such tools should also provide learning materials or tutorials and community support to easily onboard new users.

Secondly, we realize that these tools, which often provide generic feedback (see Section \ref{access}) on user interaction or their resumes, do not take into consideration the social dynamics of interviews characterized by bias. For instance, our participants reported they often had to \textit{``prove themselves''} to an employer and \textit{``convince''} them to invest in accommodations in terms of licensed ATs. This underscores the necessity for tools that go beyond generic feedback and instead offer personalized guidance that mirrors the real-life biases and obstacles BVI job seekers face. Such tools should help users develop effective communication strategies to articulate their competencies and accommodation needs confidently, without feeling like they are \textit{``imposing''} on potential employers. Tools should provide constructive feedback that helps them build a positive self-image, enabling them to demonstrate confidently in interviews that they can handle workplace collaboration without issues. By fostering a sense that they can \textit{``do things independently,''} these tools can help alleviate the emotional burden of feeling like a \textit{``weak''} candidate for needing additional support.

In addition to the above, the design of these tools must consider interaction behaviors in the absence of visual cues while providing feedback. The study underscores how BVI individuals greatly benefit from detailed verbal explanations and feedback on interactive dialogues, particularly in virtual settings. Such feedback can help them self-reflect and regulate how they handle professional social conversations with minimal pressure. For instance, AI-driven virtual interview platforms could record mock interviews and analyze verbal responses, offering insights into strengths and areas for improvement. This analysis would allow BVI users to reflect on their performance and understand how to enhance their communication and presentation skills. By providing specific, actionable feedback on aspects such as clarity, tone, and body language, these tools can help users gain a deeper understanding of how they are presented to sighted peers.

Lastly, these tools should account for the diverse socioeconomic and linguistic backgrounds of BVI individuals. Modern AI-driven systems are well-equipped to support a multilingual collaborative learning environment~\cite{wang2024large}, which can assist BVI individuals in simultaneously learning languages, improving professional conversation skills, and effectively communicating technical competence.

\subsection{Limitations and Future Work.} 
An important limitation of our study stems from the fact that we primarily relied upon self-report data drawn exclusively from BVI job seekers where we did not account for the varying levels of familiarity and access to technology among participants. Some BVI individuals may have had limited exposure to advanced assistive technologies, potentially influencing their responses and experiences.

Another limitation of our study was that we asked participants about their experiences during interviews, specifically the attitudes of interviewers towards accessibility, and reported these findings directly. However, we did not capture the specific demands of the roles for which the participants were applying, nor did we obtain the interviewers' perspectives. We recognize the importance of understanding the viewpoints of industry experts who hire people with severe vision impairments. In the future work we will focus on interviewing recruiters to identify the key skill sets and requirements they seek when hiring BVI individuals.

Additionally, participants who struggle to find jobs may develop biases that influence their job-seeking efforts, potentially affecting the quality of our data. To mitigate this, future research could involve firsthand observation of BVI individuals during interview calls to better understand the moment-to-moment interactions between them and the interviewers. This approach would provide more accurate insights into their experiences and challenges during the interview process.

Various career development and digital employment tools exist today~\cite{dillahunt2018designing,hayes2015mobile}; however, they often provide a one-size-fits-all solution and fail to address the specific needs of BVI individuals (see Section~\ref{access}). While we cannot definitively state the effectiveness of existing tools for BVI individuals, our future research will thoroughly investigate the impact of various intervention methods and technologies aimed at enhancing employability skills in job seekers similar to research work by Hendry et al.~\cite{hendry2017homeless}. Lastly, we plan to leverage the findings of this study to develop targeted intervention tools that will boost employbility among BVI job seekers. These tools will be designed to address the unique challenges faced by BVI individuals in the job market, offering tailored guidance and support to enhance their job-seeking strategies.

\section{Conclusion}
Our study highlights the complex dynamics of job-seeking behaviors among visually impaired individuals, emphasizing the significant barriers they face despite their high levels of self-efficacy and adaptability. These challenges are often exacerbated by societal stereotypes and inadequate accessibility in existing employment tools. Our findings underscore the critical need for fostering effective self-reflection, providing constructive feedback, and enhancing communication skills. Additionally, there is a pressing need to develop accessible, individualized tools to improve employability skills for BVI individuals. Addressing these specific needs can significantly improve employment outcomes and promote greater inclusion for visually impaired individuals in the job market. This study also raises new opportunities for further empirical research and design innovations aimed at supporting BVI job seekers, ultimately fostering a more inclusive job market where they can thrive and contribute meaningfully.

\begin{acks}
We sincerely appreciate the invaluable assistance of Arun Kumar Nayak, Shwetha Sandeep Rao, Shine C Simmy, and Sagar Nampally in data collection and setting up the user study.    
\end{acks}

\bibliographystyle{ACM-Reference-Format}
\bibliography{References}

\end{document}